\documentclass[aip,jcp,amsmath,amssymb,reprint,longbibliography]{revtex4-1}

\usepackage{graphicx}
\usepackage{bm} % bold math

\usepackage[utf8]{inputenc}
\usepackage[T1]{fontenc}
\usepackage{mathptmx}

\usepackage{microtype}

\usepackage[pdfusetitle,colorlinks,linkcolor=blue,citecolor=blue,urlcolor=blue]{hyperref}

\def\equationautorefname~#1\null{Eq.~(#1)\null}

\renewcommand{\Im}{\operatorname{Im}}
\renewcommand{\Re}{\operatorname{Re}}

\newcommand{\Eph}{E_{1\mathrm{ph}}}
\newcommand{\Heff}{H_{\mathrm{eff}}}
\newcommand{\Vabs}{V_{\mathrm{abs}}}

\allowdisplaybreaks{}

\begin{document}

\title{Molecular photodissociation enabled by ultrafast plasmon decay}

\author{José Torres-Sánchez}
\author{Johannes Feist}
\email{johannes.feist@uam.es}
\affiliation{Departamento de Física Teórica de la Materia Condensada and Condensed Matter Physics Center (IFIMAC), Universidad Autónoma de Madrid, E-28049 Madrid, Spain}

\date{\today}

\begin{abstract}
We propose a strategy for enabling photodissociation of a normally photostable
molecule through coupling to a nanoparticle plasmon. The large possible coupling
on the single-molecule level combined with the highly lossy nature of plasmonic
modes, with lifetimes on the order of femtoseconds, opens an ultrafast decay
channel for the molecule. For plasmon mode frequencies below the vertical
photoexcitation energy of the molecule, the difference between excitation and
emission energy is converted into vibrational energy on the molecular ground
state in a Raman-like process. Under the correct conditions, this energy can be
high enough to enable efficient photodissociation on the electronic ground
state. We demonstrate the concept using numerical simulations of the Lindblad
master equation for the hydrogen molecule in the vicinity of an aluminum
nanoparticle, and explore the photodissociation efficiency as a function of
various system parameters.
\end{abstract}

\maketitle

\section{Introduction}
Polaritonic chemistry and molecular polaritonics are fields inspired by the
possibility to manipulate and control molecular structure and dynamics through
strong interactions with confined electromagnetic modes. Interest in these
possibilities has increased strongly over the last few years from a range of
different and previously independent disciplines. General overviews over the
possibilities afforded in such systems can be found in a series of reviews
published over the last years~\cite{Ebbesen2016,Feist2018, Ruggenthaler2018,
Ribeiro2018Polariton,Hertzog2019,Herrera2020}. Most initial interest was focused
on ``macroscopic'' settings where a (large) collection of molecules interacts
with optical cavity modes (e.g., in Fabry-Perot planar mirror cavities), and in
which the so-called strong-coupling regime can be reached relatively easily with
organic molecular materials~\cite{Lidzey1999}. This regime is reached when the
(collective) interaction strength between light and matter excitations becomes
larger than the relevant dissipation rates in the system. The excitations of the
coupled system then become hybrid light-matter excitations, so-called
polaritons, that are split in energy by the so-called Rabi splitting. In
particular, the possibility to manipulate (photo)chemical reactions in such
settings has been demonstrated experimentally~\cite{Schwartz2011,Hutchison2012,
Stranius2018,Peters2019,Polak2020} and analyzed theoretically~\cite{Galego2015,
Herrera2016,Kowalewski2016Cavity,Galego2016,Galego2017,Vendrell2018Collective,
Groenhof2018,Groenhof2019}.

While collective light-matter coupling involving a macroscopic number of
molecules can be used to modify a variety of molecular properties, its influence
at the single-molecule level is limited due to the delocalized nature of
polaritons in such settings~\cite{Galego2015,Cwik2016,Groenhof2018,
Groenhof2019}. Much larger and more direct changes at the single-particle level
can be achieved by increasing the single-molecule coupling strength, which
requires increasing the confinement of the electromagnetic field modes. In order
to approach the single-molecule strong coupling limit at room temperature and
with the dipole transition strengths available in organic molecules, strongly
subwavelength field confinement is required, which in turn requires that the
real part of the dielectric permittivity of the ``cavity'' material becomes
negative~\cite{Khurgin2015}. At optical frequencies, this essentially means that
the only currently available setups with significant single-molecule coupling
strengths are based on metallic structures supporting plasmonic resonances
(i.e., oscillations of the free electrons in the metal). In such systems, strong
coupling can be reached with relatively few emitters~\cite{Zengin2015}, even
down to the single-molecule level~\cite{Chikkaraddy2016,Ojambati2019}. It has
also been shown that such setups can be used to modify molecular reactions such
as photo-oxidation~\cite{Munkhbat2018}. However, plasmonic resonances
unavoidably come with large losses due to a significant fraction of their energy
being stored in the motion of free electrons in the metal~\cite{Khurgin2015}.
The resulting resonance lifetimes are typically on the order of $10$~fs, short
compared to the time needed for most chemical reactions to take place. Very
recently, theoretical efforts have thus started to be directed towards making
virtue out of necessity, i.e., towards exploiting the ultrafast loss in
plasmonic nanocavities and nanoantennas to achieve desired
functionalities~\cite{Silva2020,Ulusoy2020,Felicetti2020,Antoniou2020,
Davidsson2020}.

Along similar lines, in this manuscript we devise a strategy for using ultrafast
decay induced by coupling to plasmonic modes to permit photodissociation in
molecules that do not normally do so upon photoexcitation. The general idea is
to induce ultrafast decay at a specific position on the excited-state potential
energy surface (PES) in such a way that the kinetic energy accumulated by the
nuclei after photoexcitation at the Franck-Condon point is sufficient to
overcome the dissociation barrier on the ground-state electronic surface after
photon loss, even when photodissociation is not energetically allowed on the
excited-state PES\@. We demonstrate the concept using the simplest neutral
molecule, hydrogen (H$_2$). It turns out that the potential energy curves for
the two lowest electronic states of H$_2$ exactly fulfill the conditions for
which the process we propose takes place efficiently. Additionally, aluminum
nanoantennas can provide plasmonic resonances with the required characteristics
(energy and decay rate). 

\section{Theory}\label{sec:theory}
\begin{figure}
  \includegraphics[width=\linewidth]{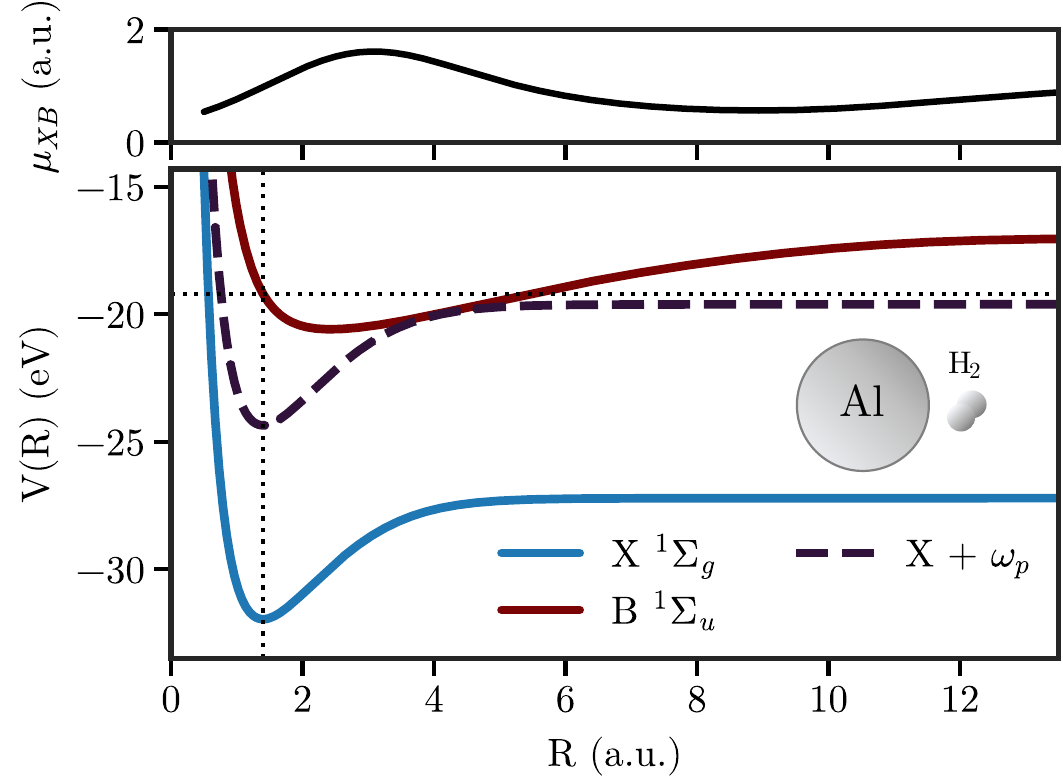}
  \caption{Main panel: Potential energy curves of the ground
  ($X$, light blue) and first excited ($B$, dark red) states of H$_2$ as a
  function of internuclear distance $R$. Also shown is the potential
  corresponding to the molecule in the ground state and a photon in the cavity
  (for $\omega_p = 7.6$~eV). Inset: Sketch of the system we consider: A H$_2$
  molecule next to an aluminum nanosphere. Upper panel: Transition dipole matrix
  element $\mu_{XB}(R)$ between $X$ and $B$ for light polarized along the
  molecular axis.}
  \label{fig:PES}
\end{figure}

The system we consider, sketched in the inset of~\autoref{fig:PES}, consists of
a single hydrogen molecule close to an aluminum nanoparticle. The molecular
electronic structure is approximated by the lowest two electronic states within
the Born-Oppenheimer approximation, the ground state $X~^1\Sigma_g^+$ and first
excited state $B~^1\Sigma_u^+$. In the following, we will refer to these states
as simply the $X$ and $B$ states, respectively. Rotational motion is neglected,
and the potential energy curves $V_X(R)$ and $V_B(R)$ as a function of the one
remaining nuclear coordinate, internuclear distance $R$, are shown in the main
panel of~\autoref{fig:PES}. The transition dipole moment between them is aligned
along the molecular axis, and its value along that axis, $\mu_{XB}(R)$, is shown
in the upper panel of~\autoref{fig:PES}. The values of $V_X(R)$, $V_B(R)$ and
$\mu_{XB}(R)$ are taken from accurate reference
calculations~\cite{Wolniewicz1993,Staszewska2002,Wolniewicz2003}. Additionally,
both the nuclear position and energy of the Franck-Condon point (corresponding
to a vertical transition from $X$ to $B$ at the minimum position in $X$,
$R_{\mathrm{FC}} \approx 1.4$~a.u.) are shown in dashed black lines. As its
energy is significantly below the dissociation limit of the $B$ state,
$V_B(R\to\infty)$, photoexcitation will typically not lead to dissociation in
the bare molecule (under instantaneous excitation, the dissociation probability
is approximately $1\%$).

The aluminum nanoparticle is modeled as a perfect sphere and represented by a
single bosonic mode describing a plasmonic pseudomode~\cite{Gonzalez-Tudela2014,
Delga2014,Delga2014a}. Such a pseudomode corresponds to a highly localized
excitation arising as a coherent superposition of many high-order multipole
modes with similar frequencies. For sufficiently short distances between the
molecule and the nanoparticle (on the order of a few nanometers and below), the
pseudomode is by far the most strongly coupled mode. When treated perturbatively
(i.e., under the assumption of weak light-matter coupling), it is responsible
for quenching~\cite{Dulkeith2002,Anger2006,Kuhn2006,Galloway2009}, but this
simple picture breaks down at short emitter-nanoparticle distances where the
coupling becomes comparable to the decay rates and the strong-coupling regime is
entered~\cite{Delga2014}. Although such pseudomodes are nonradiative and cannot
be excited (and thus detected or characterized) by far-field radiation, they can
still be used as effective quantized ``cavity'' modes with large coupling
strengths that can be useful for applications in polaritonic chemistry and
molecular polaritonics, such as photoprotection~\cite{Felicetti2020}. In order
to quantize the pseudomode we calculate the electromagnetic spectral
density~\cite{Novotny2012} $J(\omega) =
\frac{\mu^2\omega^2}{\pi\hbar\varepsilon_0 c^2} \vec{n} \cdot \Im
\mathbf{G}(\vec{r}_e,\vec{r}_e,\omega) \cdot \vec{n}$, where
$\mathbf{G}(\vec{r}_1,\vec{r}_2,\omega)$ is the dyadic Green's function solving
the (macroscopic) Maxwell equations at frequency $\omega$, $\vec{r}_e$ is the
emitter position, $\vec{n}$ is the orientation of its transition dipole moment,
and $\mu$ is its amplitude. We then use the well-known fact that Lorentzian
peaks in $J(\omega)/\mu^2$ correspond to lossy quantized modes with mode energy
$\omega_p$, loss rate $\kappa$, and single-photon field strength $\Eph$
determined by the peak position, bandwidth, and amplitude of the
peaks~\cite{Imamoglu1994,Grynberg2010,Gonzalez-Tudela2014,Delga2014,
Rousseaux2016}. While this approach can be extended to more complex spectral
densities by allowing interactions between the quantized
modes~\cite{Medina2020}, this is not necessary here as the relevant spectral
densities are dominated by a single peak.

The Hamiltonian resulting from the above considerations is ($\hbar=1$ here and
in the following)
\begin{multline}\label{eq:H}
  H = \omega_p a^\dagger a + V_X(R) + \omega_m(R) \sigma^\dagger \sigma \\
  + \Eph \mu_{XB}(R) (\sigma^\dagger a + \sigma a^\dagger),
\end{multline}
where $a$, $a^\dagger$ are the plasmon annihilation and creation operators,
$\omega_m(R) = V_B(R)-V_X(R)$ is the position-dependent molecular excitation
energy, and $\sigma = |X\rangle\langle B|$ is the molecular de-excitation
operator. We have here treated the light-matter interaction within the dipole
approximation, and additionally used the rotating wave approximation in which
the total number of excitations (molecular excitations + photons) is conserved.
This is a good approximation as long as the coupling strength is small compared
to the transition energies, i.e., the system does not enter ultrastrong
coupling~\cite{FriskKockum2019}. We also note that the pseudomode (and
nanoparticle plasmons in general) are to a very good approximation quasistatic
excitations that only interact with the molecule through the longitudinal
electric field (i.e., Coulomb interactions). This has the consequence that
irrespective of whether the quantized EM field is treated using minimal coupling
or in the Power-Zienau-Woolley picture, the light-matter interaction within the
dipole approximation is simply given by $\vec{E}\cdot\vec{\mu}$ and no dipole
self-energy term is present in the Hamiltonian~\cite{Galego2019,Feist2020}.

In the following, we will use the notation $|i,n\rangle$, with $i = (X,B)$ and
$n = (0,1,\ldots)$, for the combined electronic-photonic state. The main panel
in \autoref{fig:PES} additionally shows the potential energy surface for the
state $|X,1\rangle$, corresponding to the molecule in its ground state and a
single plasmon excitation. The associated potential energy is simply $V_X(R) +
\omega_p$, i.e., the ground-state potential energy surface shifted upwards by
that amount~\cite{Galego2015,Feist2018}. The plasmon energy $\omega_p=7.6$~eV is
chosen such that the curves corresponding to plasmonic and molecular excitation
touch tangentially (at $R\approx 4$~a.u.) while the dissociation limit of the
$|X,1\rangle$ curve lies below the Franck-Condon energy for excitation of the
bare molecule (the $|B,0\rangle$ state). This already implies that from purely
energetic considerations, the presence of the plasmonic nanoparticle can open a
photodissociation channel that does not exist in the bare molecule. In the
following, we will show that this cavity-enabled photodissociation channel
indeed is opened and will discuss its properties. Taking into account the
ultrafast decay of the plasmonic pseudomode, we have to explicitly treat
dissipation in the system in order to obtain an accurate representation of the
dynamics of the coupled system. As mentioned above, a mode corresponding to a
Lorentzian peak in the spectral density leads to dynamics described by a
Lindblad master equation~\cite{Lindblad1976,Manzano2020} for the system density
matrix $\rho$,
\begin{equation}\label{eq:master}
    \partial_t \rho = -i [H, \rho] + \kappa L_{a}[\rho],
\end{equation}
where $L_A[\rho] = A\rho A^\dagger - \frac12 \{A^\dagger A, \rho \}$ is a
\emph{Lindblad dissipator}, with $\{A,B\} = AB + BA$ the anticommutator. The
second term in $L_A[\rho]$ is responsible for probability disappearing from the
decaying state, while the first term makes it reappear in the new state that the
decay leads to. The first term is thus also variously called the ``refilling'',
``feeding'', or ``quantum jump'' term. \autoref{eq:master} can be rewritten as
\begin{equation}\label{eq:master_eff}
  \partial_t \rho = -i (\Heff \rho - \rho \Heff^\dagger) + \kappa a \rho a^\dagger
\end{equation}
where $\Heff = H - \frac{i}{2} \kappa a^\dagger a$ is an effective non-Hermitian
Hamiltonian. While we solve the full Lindblad master equation here, we mention
that in situations where the refilling term is
negligible~\cite{Saez-Blazquez2017} or the state reached by it is not of
interest~\cite{Felicetti2020}, the dynamics can equivalently be described by the
Schrödinger equation of a wave function evolving with the effective
Hamiltonian~\cite{Visser1995}, leading to significant savings in computational
effort as only a wave function instead of a density matrix has to be propagated.
However, this simplification is not applicable when the wave packet dynamics
after decay are required~\cite{Davidsson2020}, as is the case for the setup
discussed here.

We numerically discretize the nuclear coordinate $R$ using a finite element
discrete variable representation (FEDVR)~\cite{Rescigno2000,Schneider2005} in
which the domain is split into finite elements and a high-order DVR basis is
used in each element. This approach allows for high flexibility in adapting both
the element size and polynomial order to the local requirements of the problem,
while still leading to a very sparse representation of derivative operators and
potentials as block-diagonal and diagonal matrices, respectively. It typically
achieves significantly higher precision than finite difference grids for the
same computational effort. Most calculations shown below use $46$ equally sized
elements spanning from $R=0.5$~a.u.\ to $R=17$~a.u., with polynomial order $9$
inside each element, leading to $369$ nuclear basis functions. Some results are
obtained with ``better'' grids, but we have checked that the results are
visually indistinguishable from those obtained with the parameters given here,
and thus do not mention those cases explicitly. In order to prevent unphysical
reflections at the end of the grid when the molecule dissociates, we
additionally introduce a complex absorbing potential for $R>R_\mathrm{abs}$,
given by $\Vabs(R) = C (R-R_\mathrm{abs})^4$, where $R_\mathrm{abs} = 12$~a.u.\
and $C = 10^{-4}$~a.u.\ determines the strength of the absorber. In order to
maintain trace-preserving dynamics, we introduce this potential through Lindblad
dissipators to a fictitious ``dissipated'' nuclear basis function
$|R_\mathcal{D}\rangle$ that is not otherwise coupled to the system. The
absorbing potential is then described by additional Lindblad dissipators in the
master equation,
\begin{equation}\label{eq:master_abspot}
\partial_t \rho = -i [H, \rho] + \kappa L_{a}[\rho] + \sum_{i=1}^{N_r} 2 \Vabs(R_i) L_{|R_\mathcal{D}\rangle\langle R_i|}[\rho],
\end{equation}
where $|R_i\rangle$ is the FEDVR basis function corresponding to nuclear
position $R_i$. These additional dissipators add a complex absorbing potential
to the effective Hamiltonian, while also collecting the absorbed probability in
the density matrix at ``grid point'' $R_\mathcal{D}$. This approach allows
straightforward monitoring of the electronic-plasmonic potential energy curve on
which absorption (i.e., dissociation) happened.

\begin{figure}
  \includegraphics[width=\linewidth]{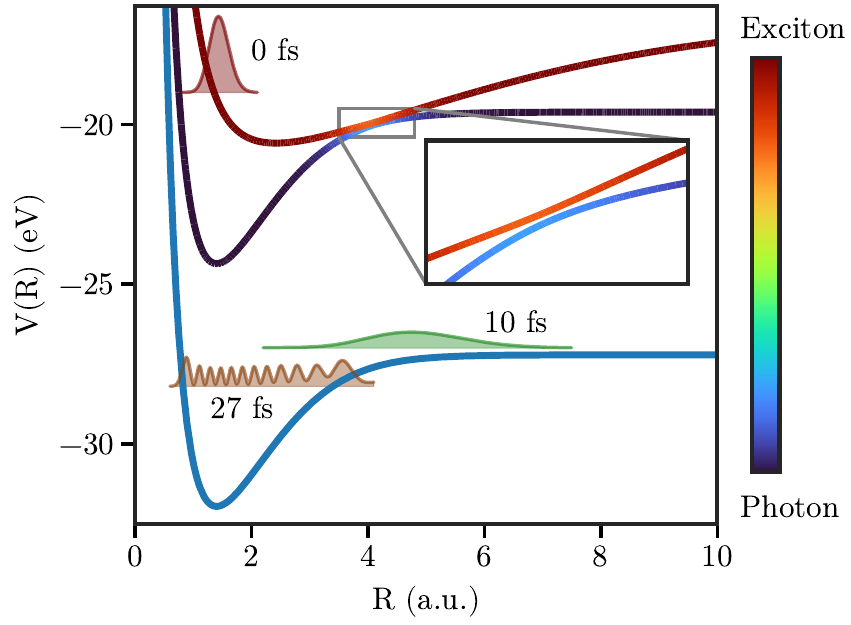}
  \caption{Polaritonic potential energy curves calculated using the effective
  non-Hermitian Hamiltonian defined in \autoref{eq:master_eff}, for a
  single-photon field strength of $\Eph=70~\mathrm{mV}/a_0$ and cavity loss rate
  $\kappa=0.476$~eV. The line color indicates the degree of hybridization. The
  zoom shows the region where the uncoupled curves become (nearly) degenerate.
  Also shown are three snapshots of the nuclear wave packet at $t=0$~fs,
  $t=10$~fs, and $t=27$~fs (the amplitude for the latter two is multiplied by a
  factor $5$ for better visibility). For $t=27$~fs, a tail corresponding to the
  dissociating part of the wavepacket at $R>4$~a.u.\ is not shown.}
  \label{fig:PoPES}
\end{figure}

For the numerical implementation, we rely on NumPy~\cite{NumPy2020} and
QuTiP~\cite{Johansson2012,Johansson2013}. Since the master equations become
relatively large, we have implemented a custom master equation solver that runs
on graphics processing units (GPU) by exploiting the
CUDA.jl~\cite{Besard2019Effective,Besard2019Rapid} and
DifferentialEquations.jl~\cite{Rackauckas2017,Rackauckas2019} packages for the
Julia language~\cite{Bezanson2017}. All plots were prepared using
matplotlib~\cite{Hunter2007,Caswell2020}.

\section{Results}
\begin{figure*}
  \includegraphics[width=\linewidth]{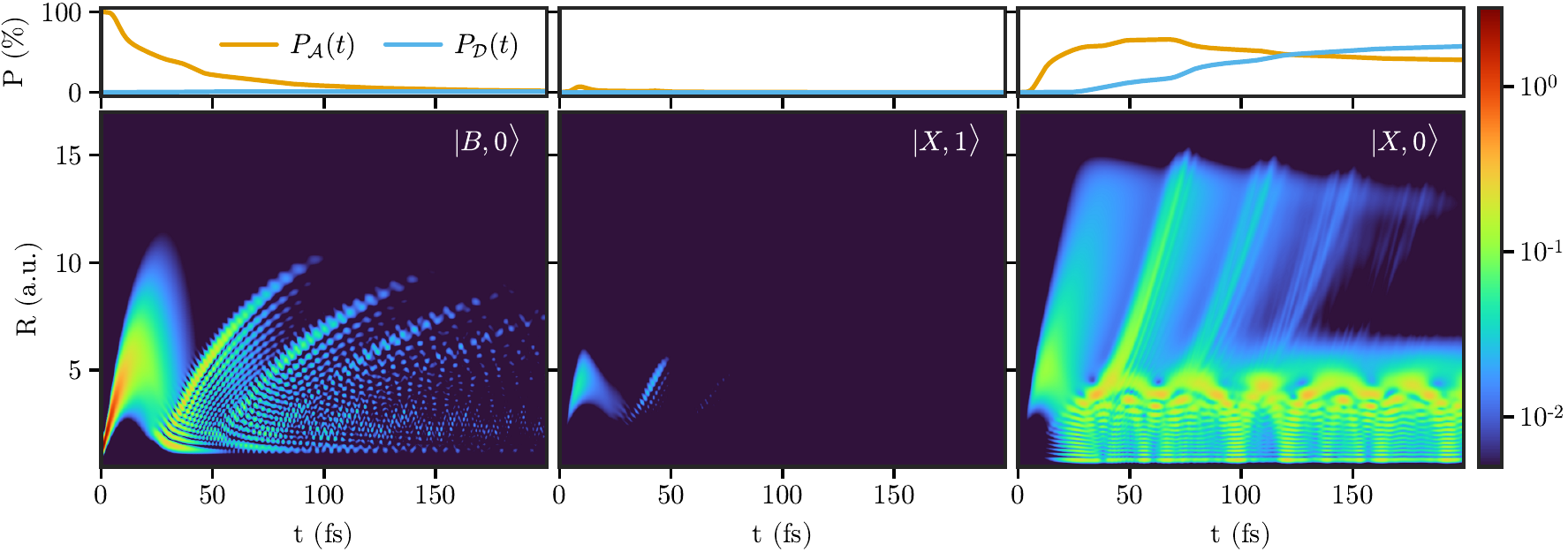}
  \caption{Lower panels: Nuclear wavepacket (probability density) corresponding
  to the excited molecule, $|B,0\rangle$, a plasmon excitation in the
  nanoparticle $|X,1\rangle$, and the ground state $|X,0\rangle$ as a function
  of time. Upper panels: Total integrated probability $P_\mathcal{A}(t)$ in the
  ``active'' nuclear region and probability $P_\mathcal{D}(t)$ of dissociation
  for each electronic-plasmonic state.}
  \label{fig:WF}
\end{figure*}

We initially set a plasmon mode frequency of $\omega_p = 7.6$~eV with decay rate
$\kappa = 0.476$~eV and quantized field strength $\Eph = 70~\mathrm{mV}/a_0$,
where $a_0$ is the Bohr radius\footnote{These uncommon units for the electric
field strength, $1~\mathrm{mV}/a_0 \approx 18.9~\mathrm{MV}/\mathrm{m}$, allow
straightforward calculation of the light-matter coupling strength $g = \Eph \mu$
in meV for a transition dipole $\mu$ given in atomic units $e a_0$, as in
\autoref{fig:PES}.}. These parameters correspond approximately to a hydrogen
molecule at a distance of slightly less than $1$~nm from an aluminum nanosphere
with a radius of $20$~nm embedded in a background dielectric material with
refractive index $n=1.75$. The corresponding polaritonic potential energy
curves~\cite{Feist2018}, calculated using the effective Hamiltonian, are shown
in \autoref{fig:PoPES}. The dynamics are initialized assuming ultrafast vertical
excitation from $X$ to $B$, i.e., with an initial state that corresponds to the
vibrational ground state on $V_X(R)$ promoted to $V_B(R)$. This initial
wavepacket is shown in \autoref{fig:PoPES} in red. The vertical excitation
energy at the equilibrium position $R_0 \approx 1.4$~a.u.\ corresponds to
$\omega_m(R_0) \approx 12.75$~eV, far detuned from the plasmon resonance. As the
wavepacket starts oscillating and spreading on the $B$ curve, within less than
$10$~fs it reaches the region where the cavity and molecular transition are on
resonance. There, the coupling to the plasmonic mode leads to population
transfer into the plasmonically excited surface, which almost immediately decays
to the ground state. As we have recently shown, the photons emitted from a
plasmonic cavity in similar setups can give direct optical access to ultrafast
nuclear motion~\cite{Silva2020}. However, in the current case, the plasmonic
mode is nonradiative and no photons are emitted into the far field. Nonetheless,
the plasmonic decay conserves the kinetic energy accumulated by the nuclear
wavepacket during its propagation on the $B$ surface~\cite{Palacios2013}. The
nuclear wavepacket arriving on the ground-state surface $V_X(R)$ (shown in green
at $t=10$~fs in \autoref{fig:PoPES}) can thus keep propagating, with a
significant fraction overcoming the dissociation barrier. At the same time, some
fraction of the wavepacket remains trapped in the bound vibrational states of
$V_X(R)$, shown in brown for $t=27$~fs in \autoref{fig:PoPES}.

To get detailed insight into the wavepacket dynamics, we show the nuclear
probability density on each of the three relevant states $|B,0\rangle$,
$|X,1\rangle$, and $|X,0\rangle$ as a function of time in \autoref{fig:WF}. This
shows the nuclear wavepacket initially oscillating on the molecular excited
state $|B,0\rangle$. The light-matter coupling then leads to efficient energy
transfer to the plasmonic state $|X,1\rangle$ at the nuclear positions where
they are close in energy, roughly between $R=3$~a.u.\ and $R=7$~a.u. The
plasmonic state almost immediately decays after excitation, such that the
wavepacket is directly transferred to the $|X,0\rangle$ state. However, it is
clearly visible in the third panel of \autoref{fig:WF} that a significant
fraction of the wavepacket that decays to the ground state overcomes the
dissociation barrier and propagates to large internuclear distances $R$, where
it is absorbed by the absorbing potential. Since the nuclear wavepacket in the
lower panels of \autoref{fig:WF} is shown in logarithmic scale, which inhibits
easy visual interpretation of absolute probabilities, in the upper panels we
additionally show the integrated probabilities in the ``active'' nuclear region
of each electronic-plasmonic state, $P_\mathcal{A}(t)$, as well as the already
dissociated component $P_\mathcal{D}(t)$. This demonstrates clearly that the
light-matter coupling through the plasmonic pseudomode induces ultrafast decay
of the nuclear wavepacket from the initial excited state, with most of the
probability disappearing within about $100$~fs. Furthermore, it also shows that
after decaying to the ground state $|X,0\rangle$, a significant portion of the
nuclear wavepacket dissociates, reaching more than $50\%$ for the present
parameters. The remaining wavepacket ends up in bound vibrational levels on the
ground-state surface.

\begin{figure*}
  \includegraphics[width=\linewidth]{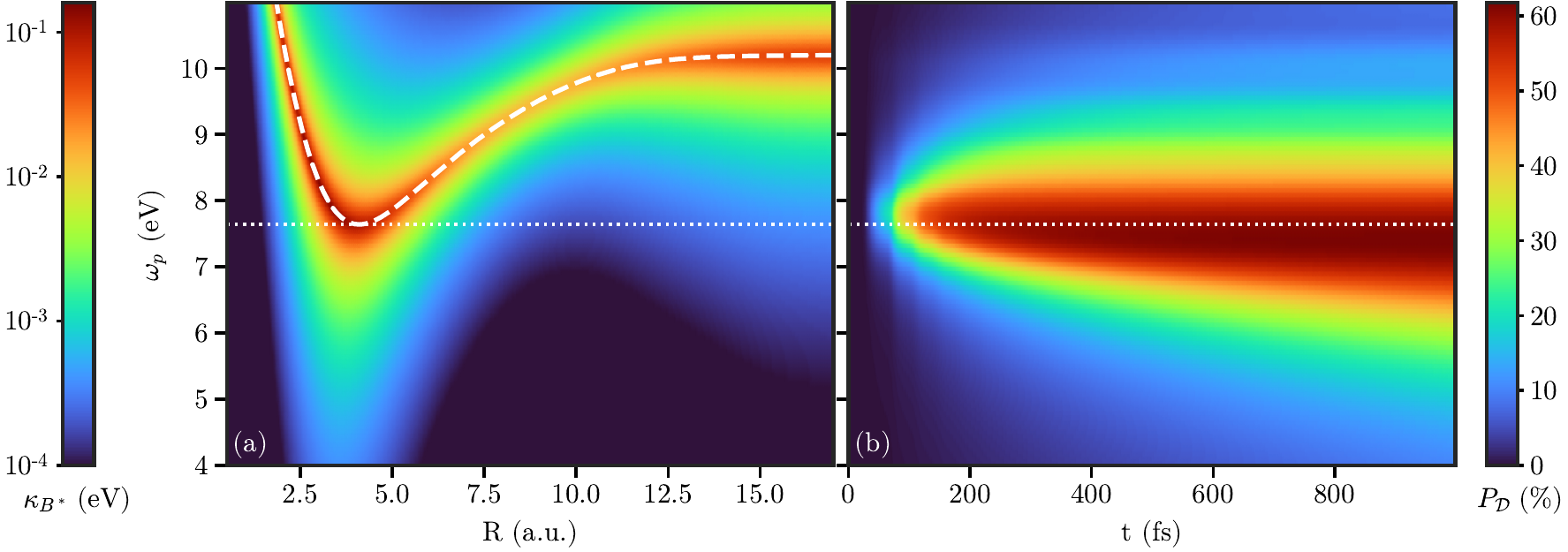}
  \caption{(a): Coupling-induced decay rate $\kappa_{B^*}(R)$ on the polaritonic
  potential energy curve corresponding to the molecular excited state as a
  function of plasmon mode frequency $\omega_p$. The dashed white line
  corresponds to the molecular excitation frequency $\omega_m(R)$. (b):
  Dissociation probability after $1000$~fs for the same conditions. The dotted
  white line in both panels marks the minimum value of $\omega_m(R)$.}
  \label{fig:time_dis}
\end{figure*}

We note that due to the highly lossy nature of aluminum, the decay rate of the
pseudomode (and other plasmonic resonances) is quite large ($\kappa =
0.476$~eV), with an associated lifetime of only $\tau = \hbar/\kappa \approx
1.38$~fs, even shorter than for the more commonly used metals such as gold or
silver, which have plasmon resonances in the visible spectrum. The strong
coupling regime is entered when $g>\kappa/4$ if the condition is taken that at
resonance $\omega_p = \omega_m(R)$, the real part of the eigenenergies of the
effective Hamiltonian starts to split, and for slightly larger $g$ if one
requires that the corresponding peaks can be spectrally
distinguished~\cite{Torma2015}. The maximum value of the transition dipole
moment $\mu_{XB}(R)$ is approximately $1.61~e\,a_0$, reached at $R=3.1$~a.u. For
realistic parameters, with $\Eph$ reaching up to about $100$~mV/$a_0$, the
system we treat will thus stay within the weak coupling regime or at best barely
enter strong coupling. However, since the coupling is quite strong even when
smaller than the decay rate, the plasmonic pseudomode serves as an effective
decay channel at internuclear distances where $\omega_m(R)$ is sufficiently
close to $\omega_p$. The ultrafast decay we see above can thus be understood to
a good approximation as an extreme Purcell effect, with the emissive lifetime
decreasing from about $\approx 0.5$~ns for the bare molecule down to the
femtosecond scale due to the pseudomode. However, it should be stressed that an
important aspect of the strategy we propose is the energy selectivity of the
plasmonic pseudomode, i.e., the fact that it does not ``just'' enhance emission,
but does so at a specific energy. Within a temporal picture, this allows to
enhance decay at nuclear positions where the initial wavepacket has already
accumulated enough kinetic energy to subsequently dissociate. Energetically, the
approach can be understood as selecting the energy difference between the
initially absorbed photon and the subsequently emitted one. This difference is
necessarily converted into vibrational energy (like in Raman scattering) which
can then be used to dissociate the molecule in its ground state. A further
important point is that this process happens on ultrafast timescales, so that,
e.g., intramolecular vibrational energy redistribution (which would only occur
in more complex molecules with multiple nuclear degrees of freedom) does not
have time to dissipate the accumulated kinetic energy between different modes.

We thus next explore the energy sensitivity of the loss-induced
photodissociation process in more detail. To do so, we scan the frequency of the
plasmonic pseudomode $\omega_p$ over a large interval (from $4$ to $11$~eV)
while keeping all other parameters fixed. Since for the given coupling strength,
the system is mostly in the weak coupling regime, a useful quantity to plot is
the induced effective decay rate of the (cavity-modified) excited molecular
state $B^*$. Here, the asterisk denotes polaritonic states obtained by
diagonalizing the effective Hamiltonian $\Heff$ within the Born-Oppenheimer
approximation (i.e., separately for each $R$), and $B^*$ is the state that is
closest in character to the bare molecular excited state $B$. In the effective
Hamiltonian, the energies (and thus the polaritonic PES) become complex, and the
local decay rates are given by $\kappa_{B^*}(R) = -2\Im V_{B^*}(R)$. Within the
(almost) weak coupling regime considered here, the modifications of the real
part of the surface are much smaller, $\Re V_{B^*}(R) \approx V_B(R)$. We thus
show $\kappa_{B^*}(R)$ as a function of plasmon frequency $\omega_p$ in
\autoref{fig:time_dis}(a). The white dashed line in the figure is given by
$\omega_m(R)$, i.e., the position-dependent molecular excitation energy. The
plasmon-induced molecular decay rate is most strongly enhanced when the plasmon
is on resonance with the molecular transition, $\omega_p\approx \omega_m(R)$, as
could be expected and has been previously proposed as a tool for probing
molecular dynamics by ultrafast emission~\cite{Silva2020}. The cavity-induced
decay rate for non-zero detuning depends on both the detuning and the coupling
strength. For $g(R) = \Eph\mu_{XB}(R) \ll \kappa/4$, it can be well approximated
by a Lorentzian function, $\kappa_{B^*}(R) \approx \frac{4 g(R)^2}{4 \delta(R)^2
+ \kappa^2}$, where $\delta(R) = \omega_m(R) - \omega_p$.

We now focus on how this frequency dependence of the pseudomode-induced decay
affects the photodissocation process. To that end, \autoref{fig:time_dis}(b)
shows the total dissociation probability as a function of time for the same
range of plasmon frequencies. We first note that at short times, there is a
step-wise behavior that stems from the initially localized character of the
nuclear wavepacket, which creates a burst of dissociation every time it passes
the nuclear distances where decay is efficient~\cite{Silva2020}, and then
proceeds towards dissociation. At later times, the nuclear wavepacket
delocalizes due to the non-harmonic character of the involved potentials, and
the step-wise behavior is washed out.

As a function of frequency, there is a dominant peak centered just below
$\omega_p = \min\omega_m(R)$, i.e., the minimum of the molecular transition
frequency (indicated by a thin dotted white line in the figure). This reinforces
that, as discussed above, the molecular decay rate shown in panel (a) does not
by itself determine the dissociation probability. In addition to inducing
ultrafast decay on the excited-state surface, the decay has to happen after the
wavepacket acquires enough vibrational energy to dissociate after reaching the
ground state. This consideration implies that smaller $\omega_p$ should be
favored, as less of the total available energy is then spent on creating the
plasmon excitation, and the remainder can be used for overcoming the
dissociation barrier. Combining this with the competing requirement of being
close to resonance to achieve efficient decay explains the results observed
here. However, it should additionally be noted that while at the maximum times
considered here, both very low and very high plasmon frequencies appear
similarly efficient in inducing dissociation, their long-time behavior is
markedly different: While at high frequencies, the wavepacket decays efficiently
(at least as long as the cavity frequency is not significantly above the
molecular excitation frequency at the Franck-Condon point,
$\omega_m(R_{\mathrm{FC}}) \approx 12.75$~eV, higher than the frequencies
considered here), it primarily does so with little vibrational energy and thus
stays in the bound vibrational states on the electronic ground state curve
$V_X(R)$. In contrast, the induced decay slows down when the plasmon frequency
is below the minimum of $\omega_m(R)$, but the wave packet after decay acquires
considerable vibrational energy and can thus dissociate with high probability.
The end result is that at high frequencies, the dissociation probability is
already essentially saturated for the latest time shown in
\autoref{fig:time_dis}(b), $t_f = 1000$~fs, while at low frequencies it keeps
rising even at longer times. For the lowest frequency considered, $\omega_p =
4$~eV, the dissociation probability rises to over $45\%$ when propagation is
performed until $t=10$~ps, and fitting to an exponential saturation curve
implies that the limiting value is $P_\mathrm{D}(t\to\infty) \approx 49\%$. This
implies that the plasmon-induced photodissociation efficiency at lower plasmon
frequencies can also be significant, but the reduced speed means that this case
is more sensitive to effects not considered here (such as collisions or
intramolecular vibrational energy redistribution in more complex molecules).

\begin{figure}
  \includegraphics[width=\linewidth]{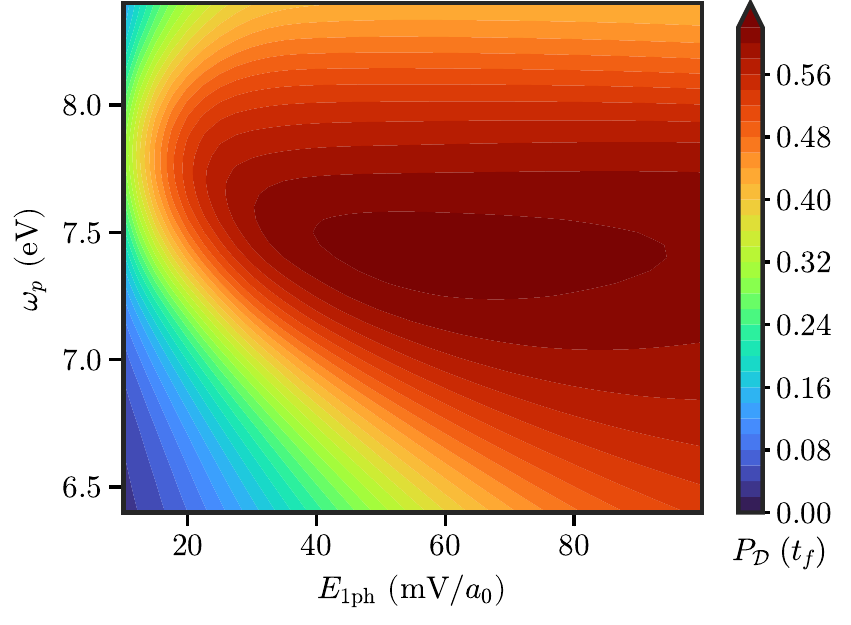}
  \caption{Dissociation probability $P_\mathcal{D}(t_f)$ at $t_f = 1000$~fs as a
  function of single-photon field strength $\Eph$ and plasmon mode frequency
  $\omega_p$.}
  \label{fig:diswf_scan} 
\end{figure}

We finally check how strongly the effects studied here depend on the coupling
strength to the plasmon mode, which is strongly dependent on the distance to the
nanoparticle. The total dissociation probability at time $t_f = 1000$~fs is
shown as a function of plasmon frequency $\omega_p$ and quantized single-photon
field strength $\Eph$ in \autoref{fig:diswf_scan}. The considered values for
$\Eph$, ranging from $10$~mV/$a_0$ to $100$~mV/$a_0$, correspond to distances of
roughly $0.5$~nm to $2$~nm, which lies within the experimentally accessible
range for nanoplasmonic antennas~\cite{Chikkaraddy2016,Ojambati2019}. The
effective mode volume, $V_{\mathrm{eff}} = \hbar\omega_p/(2\varepsilon_0
\Eph^2)$ for the largest couplings considered here is on the order of
$20$~nm$^3$. While we stay within a single-mode description here, it should be
noted that for increasing distances, the pseudomode stops being the most
strongly coupled mode, and the dipole mode (which is radiative and has lower
frequency) becomes dominant~\cite{Delga2014}. \autoref{fig:diswf_scan} shows
that, as could be expected, increasing coupling strengths lead to increasing
dissociation efficiency. It can also be appreciated that the frequency range
over which dissociation is efficient increases with increasing field confinement
(i.e., larger $\Eph$). However, we also find that there is an optimal value for
the coupling strength, i.e., the photodissociation probability starts dropping
again as the coupling strength increases above an optimal value. Consistent with
results found in the photoprotection efficiency of uracil
molecules~\cite{Felicetti2020}, the optimal value is close to the onset of the
strong-coupling regime. While the situation in a molecule with nuclear motion is
somewhat more complicated than in two-level emitters, the underlying reason for
this can be understood by taking into account that plasmon-induced decay is most
efficient just at the limit between weak and strong
coupling~\cite{Bozhevolnyi2016}, where the effective Hamiltonian has an
exceptional point~\cite{El-Ganainy2018}.

\section{Summary}
To summarize, we have theoretically investigated a strategy for using highly
lossy plasmonic modes to enable and enhance photodissociation in molecules for
which the excited-state surface reached by photoexcitation is non-dissociative.
We used the H$_2$ molecule next to an aluminum nanoparticle as a model system.
For short distances, the extremely localized and nonradiative plasmonic
pseudomode then functions as an effective ``cavity'' mode with extreme
sub-wavelength field confinement. The basic idea of the strategy we proposed is
to use a resonantly enhanced Raman-like process where part of the energy
absorbed from an exciting photon is resonantly emitted into a lower-energy
plasmon resonance, with the remaining energy converted to vibrational excitation
on the molecular ground state. If this energy is high enough, the nuclear
wavepacket can then overcome the dissociation barrier. Making use of the
unavoidable high losses of plasmonic nanoantenna modes, the induced ultrafast
decay rates ensure that the whole process takes place on femtosecond scales,
faster than other relevant processes leading to energy dissipation.

We have studied in detail how the process depends on the properties of the
plasmonic modes, in particular their frequency and the achieved field
confinement, and have found that there are competing effects leading to a
pronounced maximum as a function of cavity frequency: On the one hand, lower
plasmon frequencies leave more of the initial energy available as vibrational
energy for the wavepacket after decay, while on the other hand, resonance
between the plasmon and molecular excitation enhances the overall decay rate on
the plasmon-modified potential energy curve. The competition between these two
effects leads to a maximum of dissociation efficiency close to the minimum
excitation energy of the molecule, i.e., at the point where the ground- and
excited-state surfaces are closest to each other.

The strategy we have presented adds to the growing repertoire of loss-induced
phenomena in molecular polaritonics and polaritonic chemistry~\cite{Silva2020,
Ulusoy2020,Felicetti2020,Antoniou2020,Davidsson2020}. This field has enjoyed
increasing interest recently as a way to make virtue out of necessity in
nanoplasmonic systems, where extreme subwavelength confinement can provide
coupling strengths close to or within the single-molecule strong-coupling regime
at room temperature, but losses on the femtosecond scale are unavoidable.

\begin{acknowledgments}
We thank Alicia Palacios for insightful discussions and for providing the PES of
the hydrogen molecule in digitalized form. This work has been funded by the
European Research Council through grant ERC-2016-StG-714870, and by the Spanish
Ministry for Science, Innovation, and Universities -- Agencia Estatal de
Investigación through grants RTI2018-099737-B-I00, PCI2018-093145 (through the
QuantERA program of the European Commission), and CEX2018-000805-M (through the
María de Maeztu program for Units of Excellence in R\&D).
\end{acknowledgments}

\section*{Data availability statement}
The data that support the findings of this study are available from the
corresponding author upon reasonable request.

\bibliography{references}

\end{document}